\documentclass[aps,pre,twocolumn,floatfix,superscriptaddress,showpacs]{revtex4-1}
\usepackage{amsfonts}
\usepackage{amsmath}
\usepackage{amssymb}
\usepackage{graphicx}
\usepackage[english]{babel}
\usepackage{bm}

\newcommand{\ket}[1]{|#1\rangle}
\newcommand{\bra}[1]{\langle #1|}

\begin{document}

\title{Quantum Fuel with Multilevel Atomic Coherence for \\Ultrahigh Specific Work in a Photonic Carnot Engine}

\author{Deniz T\"{u}rkpen\c{c}e}
\affiliation{Department of Physics, Ko\c{c} University, \.{I}stanbul,Sar{\i}yer 34450, Turkey}

\author{\"{O}zg\"{u}r E. M\"{u}stecapl{\i}o\u{g}lu}
\email{dturkpence@ku.edu.tr}
\affiliation{Department of Physics, Ko\c{c} University, \.{I}stanbul,Sar{\i}yer 34450, Turkey}

\date{\today}

\begin{abstract}
We investigate scaling of work and efficiency of a photonic Carnot engine with the number of quantum coherent resources. Specifically, we consider a generalization of the ``phaseonium fuel'' for the photonic Carnot engine, which was first introduced as a three-level atom with two lower states in a quantum coherent superposition by [M. O. Scully, M. Suhail Zubairy, G. S. Agarwal, and H. Walther, Science {\bf 299}, 862 (2003)],  to the case of $N+1$ level atoms with $N$ coherent lower levels. We take into account atomic relaxation and dephasing as well as the cavity loss and derive a coarse grained master equation to evaluate the work and efficiency, analytically. Analytical results are verified by microscopic numerical examination of the thermalization dynamics. We find that efficiency and work scale quadratically with the number of quantum coherent levels. Quantum coherence boost to the specific energy (work output per unit mass of the resource) is a profound fundamental difference of quantum fuel from classical resources. We consider typical modern resonator set ups and conclude that multilevel phaseonium fuel can be utilized to overcome the decoherence in available systems. Preparation of the atomic coherences and the associated cost of coherence are analyzed and the engine operation within the bounds of the second law is verified. Our results bring the photonic Carnot engines much closer to the capabilities of current resonator technologies. 

\end{abstract}

\maketitle
\section{Introduction}

A practical figure of merit to compare fuel and battery materials is the specific energy, or energy to mass ratio    
 ~\cite{chalk_key_2006,chau_overview_1999,yang_new_2010}. As a material constant, it measures the energy that will be harvested by using a unit mass of the material. About a decade ago, a highly non-traditional fuel, called ``phaseonium'', which is a three level atom with  two lower states in a quantum coherent superposition, was proposed to be used in a Photonic Carnot engine (PCE)~ \cite{scully_extracting_2003}. Phaseonium engine could work with a single heat bath and a phaseonium reservoir~\cite{scully_extracting_2003,scully_extracting_2002,Rostovtsev_extracting_2003,zubairy_coherence_2002}. This proposal stimulated much interest to quantum heat engines ~\cite{kieu_second_2004,quan_quantum_2007,allahverdyan_work_2008,
 johal_quantum_2009,wang_efficiency_2012,scully_quantum_2011,li_negentropy_2013,zhuang_quantum_2014,altintas_quantum_2014,
 Rabi_Otto_2015,uzdin_equivalence_2015,niedenzu_efficiency_2015,Superradiant_Otto_2015}.
It was later argued that existing resonator systems can not implement such an engine, due to high cavity losses and atomic dephasing \cite{quan_quantum-classical_2006}.
Here, we address a fundamental question of how the specific energy of phaseonium fuel is scaled with the number of quantum coherent levels. A favorable scaling law against decoherence and dephasing could
bring the phaseonium engine closer to available practical systems.

\begin{figure}[htb!]
\centering
\includegraphics[width=3.4in]{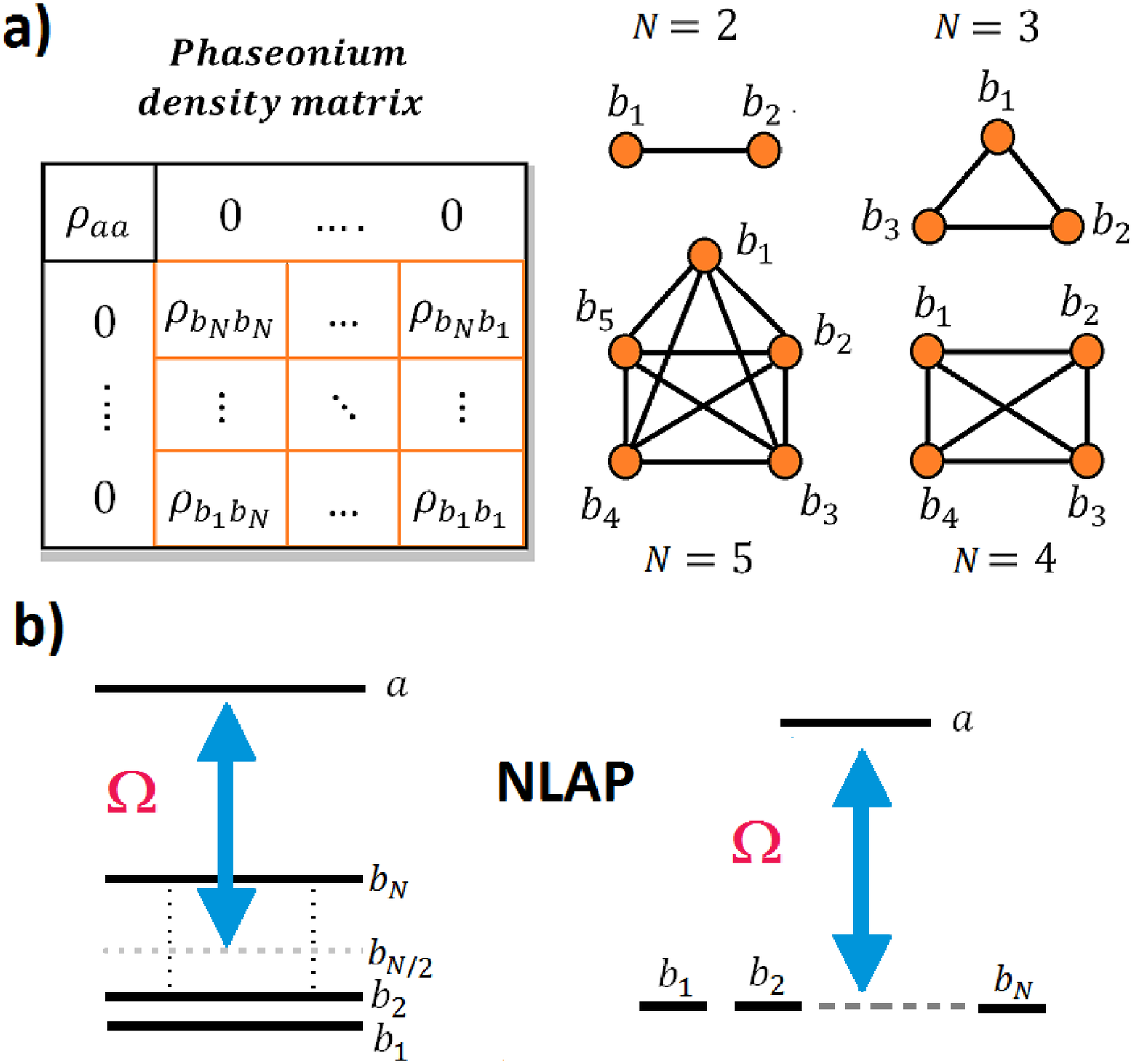}
\caption{\label{Fig1}(Color online) $N+1$ level atom phaseonium (NLAP) fuel. a) Density matrix $\rho$ and complete graph representations of NLAP. $\rho$ is  $N+1$ dimensional square matrix. Its coherent block can be represented by a complete graph with $N$ nodes and $S=N(N-1)/2$ links. Graphs were shown up to $N=5$ number of nodes. b) NLAP for non-degenerate and degenerate atoms. The excited state is denoted by $a$ and the lower levels are denoted by $b_i$ with $i=1..N$. The upper level is well separated from the lower levels by an energy $\hbar\Omega$ measured from the central lower level $b_{N/2}$.} 
\end{figure}

 We describe multilevel generalization of phaseonium fuel in Fig.~\ref{Fig1}. The block diagonal density matrix $\rho$  of an $N+1$ level atom is shown in Fig.~\ref{Fig1}(a). The excited level, denoted by ``$a$'', and the lower levels, denoted by ``$b_1, b_2,.., b_N$'' are well separated
from each other by an energy $\Omega$ measured from the central lower level $b_{N/2}$ as shown
in Fig.~\ref{Fig1}(b). The lower levels can be degenerate or non-degenerate. The diagonal elements $\rho_{aa}$, and $\rho_{bb}$, with $b \in \{b_1, b_2,..,b_N\}$, determine the level populations, while the off diagonal elements $\rho_{bb'}$, with $b'\neq b$, indicate the coherence between the levels. Coherence can be characterized by the magnitude and phase of the complex number $\rho_{bb'}$. 

Though both the amplitude and the phase of coherent superposition states can be controlled in experiments \cite{vewinger_amplitude_2007}, the main control variable for the photonic Carnot engine is the phase of the coherence as the amplitude is required to be small enough to keep the system only slightly out of thermal equilibrium. The complete graphs in Fig.~\ref{Fig1}(a) have $N$ nodes and $N(N-1)/2$ links, representing the atomic energy levels and the coherences between them, respectively. The simplest graph has $N = 2$ nodes, which is the case of the original phaseonium proposal \cite{scully_extracting_2003}. The interplay between quantum coherence and energy discussed in photon Carnot engine \cite{scully_extracting_2003} revealed that the energy content of the phaseonium with $N = 2$ can be optimized at a certain phase of the coherence. We could envision as if
we are considering more complex, larger, phaseonium molecules with the graphs having $N > 2$, corresponding to $N+1$ level atom
phaseonium (NLAP).


We can imagine different phaseonium molecules can be possible for a given atom of unit mass and explore how the specific energy of the atom depends on the size of the phaseonium molecule characterized by $N$.  Next to the phase of coherence, $N$ becomes another control parameter which could favourably contribute to the enhancement of the specific energy of the single atom quantum fuel. For $N \gg$ 1, the number of coherences would scale quadratically, $\sim N^2$. If the quadratic coherence scaling could be translated into the energy content of the atomic fuel, we could overcome the cavity losses for implementation and boost the performance of quantum Carnot engine for applications. From fundamental point of view, such scaling analysis could reveal profound difference of quantum fuel from a classical resource as such a scaling cannot exist without quantum coherence. Complete graphs of phaseonium molecules serve more than a simple counting of coherences. They emphasize the generality of our question we address in the present contribution. Can we beat decoherence with the scaling advantage of quantum coherent resources?


\begin{figure}[htb!]
 \centering
\includegraphics[width=3.2in]{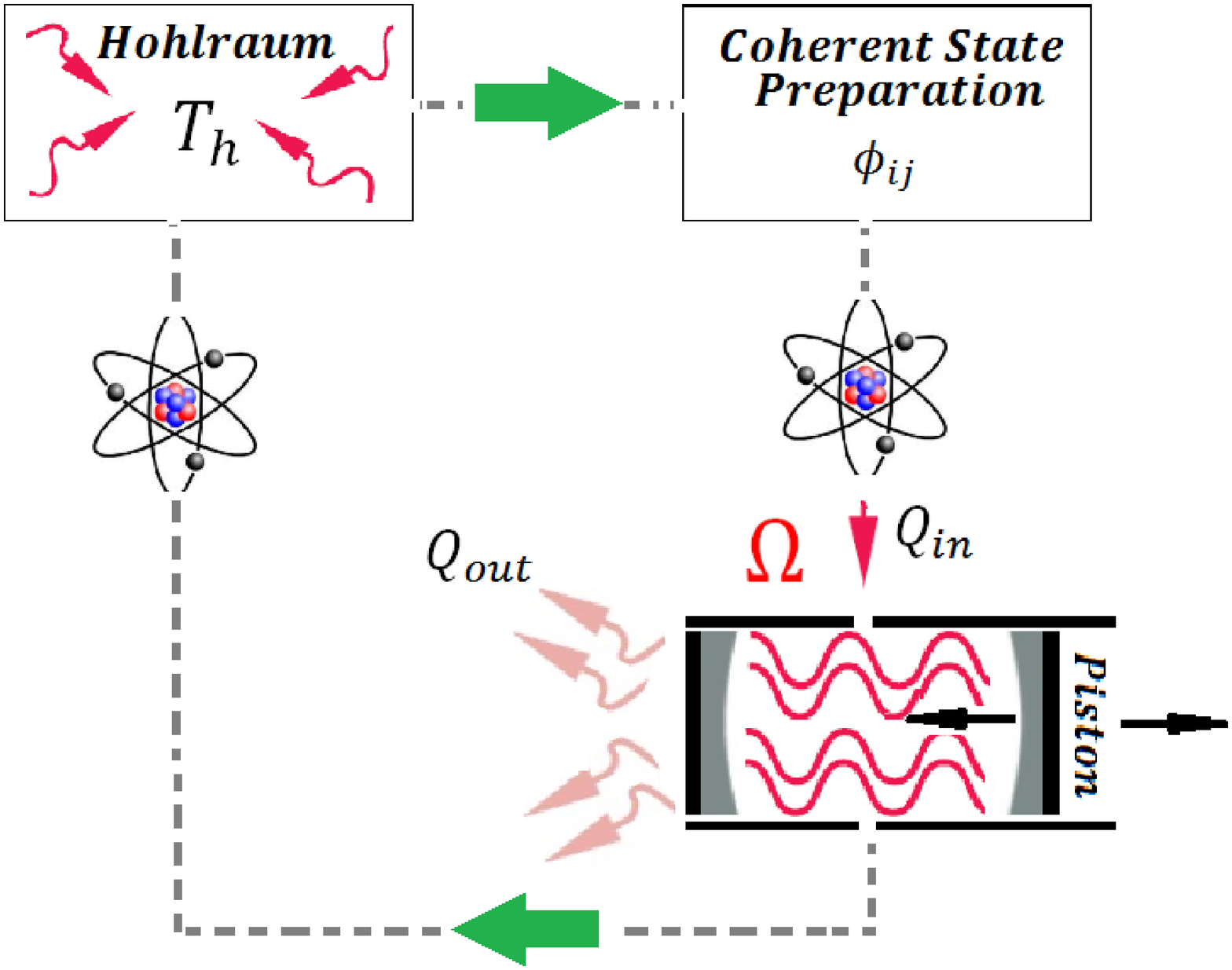}
\caption{\label{Fig2}(Color online) Photonic Carnot engine with $N+1$ level atom phaseonium (NLAP) fuel. Photon gas in a high quality cavity of frequency $\Omega$ is the working substance and the mirrors of the cavity play the role of the piston.  NLAP leaves the hohlraum at temperature $T_h$ and is subsequently prepared in a state with quantum coherence among its lower levels characterized by $N(N-1)/2$ phase parameters $\phi_{ij}$ with $i,j=1...N$. Created NLAPs are repeatedly injected into the cavity at a rate $r$ in the quantum isothermal expansion process, where heat $Q_{\mathrm{in}}$ is transferred to the cavity. The cycle continues with quantum adiabatic expansion and quantum isothermal compression and is completed with a quantum adiabatic compression. An amount of heat $Q_{\mathrm{out}}$ is rejected into the entropy sink in the isothermal compression.} 

\end{figure}  


We have recently numerically examined a superradiant quantum Otto engine~\cite{Superradiant_Otto_2015} which exhibits a similar scaling law of work with the number of atoms; though the efficiency is independent of the temperature and the number of atoms. Our present results are fully analytical, and for a Carnot cycle which can harvest work from a single heat bath and
the scaling laws with the number of coherences are both for the work and the efficiency. Furthermore, specific energy of a superradiant cluster increase linearly with the number of atoms, while here for the single atom phaseonium fuel it is quadratic. Quantum advantage in the charging power of quantum batteries with the number of qubits has been examined very recently~\cite{binder_quantacell:_2015}. Linear scaling of work with the number of qubits is reported; while due to a quantum speed up of the operation time, the charging power scales quadratically~\cite{binder_quantacell:_2015}. The preparation of phaseonium fuel and using it in PCE can be compared to charging and discharging a single qubit quantum battery with multiple quantum coherences. Phaseonium fuel or battery allows for quadratic scaling in harvested work, efficiency, and the specific energy with the number of quantum coherences. We examined the charging or preparation cost of the phaseonium battery and compared it with the harvested work by PCE. Our result verified that the second law is obeyed in our system.

\section{NLAP model and system dynamics}\label{Dynamics}


The operation of photonic Carnot engine is described in Fig.~\ref{Fig2}. The working fluid of the engine is the photon gas
in a high quality single mode cavity of frequency $\Omega$. The radiation pressure by the cavity photons applies on the cavity mirrors playing the role of the piston of the engine. The quantum fuel of the engine is an NLAP. The quantum Carnot cycle consists of two quantum isothermal and two quantum adiabatic processes.

In the isothermal expansion, NLAPs are generated and injected into the cavity at a rate $r$. The
interaction time $\tau$ between an NLAP and the cavity field is short, $\tau<1/r$, so that only one NLAP can be present in the cavity~\cite{filipowicz_theory_1986}. Coherences in NLAP are characterized by $N(N-1)/2$ phase parameters $\phi_{ij}$, with $i,j=1,2,...,N$. Coherent superposition states in $N+1$ level atom system can be generated by stimulated Raman adiabatic passage~\cite{amniat-talab_superposition_2011,bevilacqua_implementation_2013}, Morris-Shore transformation~\cite{saadati-niari_creation_2014}, or quantum Householder reflection techniques~\cite{ivanov_engineering_2006,ivanov_navigation_2007}. Thermalization of the single atom can be considered relatively fast and hence the injection rate would be limited by the time of coherence preparation. The choice of specific technique of coherence induction depends on the details of a particular implementation. If the amplitudes of the coherences are much smaller than the level populations, then NLAP can be assumed in an approximate thermal equilibrium with a thermal reservoir (hohlraum) at a temperature $T_h$. During the interaction, the mean number of photons, $\bar n$, and the cavity temperature increases; while the expansion cools down the cavity when there is no atom inside. Repeated injection of NLAPs into the cavity maintains the cavity field at a temperature $T_\phi$ by transferring a total amount of heat into the cavity as $Q_{\mathrm{in}}$. 
$T_\phi$ is an effective temperature defined in terms of the steady state photon number $\bar{n}_{\phi}$ as
 $T_{\phi}=\hbar\Omega/k\ln(1+1/\bar{n}_{\phi})$, with $k$ is the Boltzman constant. It can be higher than $T_h$ in the presence of coherence~\cite{scully_extracting_2003}. The cavity volume, and hence the frequency, change negligibly, $\Delta\Omega\ll\Omega$. 
  
The cycle continues with an adiabatic expansion where the entropy remains constant and the temperature drops as the 
$\Omega$ changes appreciably. Following step is the isothermal compression in which heat $Q_{\mathrm{out}}$ is transferred from cavity to a cold reservoir at a temperature $T_c$. The cycle is completed by adiabatic compression where the temperature is raised back to $T_\phi$. 

The net work extracted from the cycle is $W_{\mathrm{net}}=Q_{\mathrm{in}}-Q_{\mathrm{out}}$ where 
$Q_{\mathrm{in}}=T_{\phi}(S_2-S_1)$ and $Q_{\mathrm{out}}=T_c(S_3-S_4)$. The mean photon number $\bar{n}_i$ and the temperature
$T_i$ at the beginning of the $i^{th}$ stage determine the entropy $S_i$ by $S_i=k\ln(\bar{n}_i+1)+\hbar\Omega\bar{n}_i/T_i$. Using $S_1=S_4$, $S_2=S_3$, $T_1=T_2=T_{\phi}$, and $T_3=T_4=T_c$, we write $W_{\mathrm{net}}=(T_{\phi}-T_c)(S_2-S_1)$. The efficiency  of the engine is defined as $\eta=W_{\mathrm{net}}/Q_{\mathrm{in}}$. It reduces to  $\eta=1-T_h/T_{\phi}$. This coincides with the standard definition of thermodynamic efficiency in Carnot cycle and used in the original proposal of the phaseonium fuel~\cite{scully_extracting_2003,scully_extracting_2002,Rostovtsev_extracting_2003,zubairy_coherence_2002} as well as in the
arguments against its feasibility in the presence of decoherence channels~\cite{quan_quantum-classical_2006}. 
In order to present results comparable to the previous works, we calculate the efficiency as defined by these studies. To avoid any misleading impressions however, we emphasize that in practical considerations round trip efficiency can be more relevant figure of merit. The round trip efficiency of the engine should include the 
cost of the preparation of the quantum coherent atom; which would ensure the validity of the second law~\cite{zubairy_coherence_2002}. On the other hand, it was noted that the cost of quantum fuel can be expensive~\cite{scully_extracting_2002}, 
but it is still appealing as it can be used to harvest work from a single heat reservoir. 
Our objective here is not to discuss if such PCEs can be efficient enough for certain applications, but to examine if 
such devices, proposed in~\cite{scully_extracting_2003}, can produce positive work in the presence of decoherence by exploiting
a scaling advantage of multiple coherence resources, in contrast to the negative conclusions of earlier studies~\cite{quan_quantum-classical_2006}. 

During the adiabatic process $\bar{n}$ does not change
so that $\bar{n}_1=\bar{n}_4=(\exp(\hbar\Omega/kT_{c})-1)^{-1}$ and $\bar{n}_2=\bar{n}_{\phi}$. These relations reveal that work and efficiency of the photonic Carnot engine can be calculated by determining the $\bar{n}_{\phi}$ at the end of the isothermal expansion stage. 

This manuscript organised as follws: In Sec. II, we describe the N-level phaseonium model, and PCE system dynamics. We review and discuss the analytical and numerical verification of the analytical results in Sec. III. We also estimate the preparation cost of NLAP in this section. We conclude the results in Sec. IV.


\section{Results and discussions}\label{Results}


In order to find the  $\bar{n}_{\phi}$, we solve $\dot{\bar{n}}_{\phi}=\sum_{n}n\dot{\rho}_{nn}=0$ where 
$\dot{\rho}_{nn}=\bra{n}\dot{\rho}\ket{n}$.  Here $|n\rangle$ is the Fock number state for the cavity photons and $\rho$ is the reduced density matrix of the cavity field. The equation of motion for $\rho$ can be obtained by tracing the equation of motion of the complete system over atomic degrees of freedom
\begin{equation}
\dot{\rho}_{nn}=-\frac{i}{\hbar}\sum_{k}(\mbox{Tr\scriptsize at}[H^k,\rho^k]_{nn})\label{Eq.ofMot},
\end{equation}
where $H^k=H_0+H_I^k$  is the Hamiltonian of the arbitrary $k^{th}$ atom in the interaction picture relative to the cavity photons, with $H_0=\hbar\omega_a\ket{a}\bra{a}+\hbar\sum_{i=1}^N \omega_{b_i}\ket{b_i}\bra{b_i}$ and 
$H_I^k=\hbar g\sum_{i=1}^N \ket{a}\bra{b_i}\hat{a} e^{-i\Omega t}+H.c.$. Here $ \hbar\omega_a $, $ \hbar\omega_{b_i} $ are 
the energies of atomic states $|a\rangle$ and $|b_i\rangle$, with $i=1..N$, $g$ is the coupling rate between the atom and the field, 
and $\hat{a}$ is the photon annihilation operator. The model Hamiltonian describes a situation where $N+1$ level atom is coupled to a single mode cavity in a fan shaped transition scheme. A more realistic model requires consideration of multimode cavity coupled to an atom with multiple upper and lower hyperfine levels~\cite{birnbaum_photon_2005,arnold_collective_2011}. Such models can be reduced to effective single mode cavity and multilevel atom interactions~\cite{birnbaum_photon_2005} or can be directly described by generalized master equations of micromasers~\cite{reshetov_polarization_2004}. Atoms with fan shaped degenerate level schemes are also studied from the perspective of generating large superposition states~\cite{kyoseva_coherent_2006,amniat-talab_superposition_2011}. The central question for us here is the dependence of work and efficiency on the number of the superposed quantum states and we will only consider single upper level and a set of degenerate or non-degenerate lower levels for simplicity. 

Analytically calculating the right hand side of the Eq.~(\ref{Eq.ofMot}), we find (see Appendix for details)


\begin{eqnarray}\label{Eq.Mot2}
&\dot{\rho}_{nn}=-rg^2\{K_a\rho_{aa}[(n+1)\rho_{nn}-n\rho_{n-1,n-1}]\nonumber\\
&+(\sum_{i=1}^N K_{b_i}\rho_{b_ib_i}+\sum_{i<j} K_{ij}^{\phi_{ij}}|\rho_{b_ib_j}|)\nonumber\\
&\times [n\rho_{nn}-(n+1)\rho_{n+1,n+1}]\},
\end{eqnarray}
where the coefficients  $K_a,K_{b_i}$ and $K_{ij}^{\phi_{ij}}$ depend on the atomic relaxation rate $\gamma$, atomic
dephasing rate $\gamma_{\phi}$, detuning parameter $\Delta_i=\omega_{ab_i}-\Omega$, with  $\omega_{ab_i}=\omega_a-\omega_{b_i}$, and $\omega_{b_ib_j}=\omega_j-\omega_i$, as well as the coherence
parameters $\phi_{ij}$ and $|\rho_{b_ib_j}^0|$, by the relations given in the Appendix. Thus, we obtain the rate of 
change of average photon number
\begin{eqnarray}
\dot{\bar{n}}_{\phi}&=rg^2\{K_a\rho_{aa}(\bar{n}_{\phi}+1)-(R_{g_0}+R_{g_c})\bar{n}_{\phi}\}, \label{RateEq.}
\end{eqnarray}
where $R_{g_0}=\sum_{i=1}^N K_{b_i}\rho_{b_ib_i}$ and $ R_{g_c}=\sum_{i<j} K_{ij}^{\phi_{ij}}|\rho_{b_ib_j}|$.

The equation of motion for the evolution of
population elements in the density matrix coincides with the thermalization dynamics of a resonator coupled to a heat bath. 
Accordingly, the coarse grained dynamics effectively describes sequence of NLAP injected into the resonator as a mesoscopic
ensemble of $N+1$ level atoms acting as a heat bath. The off diagonal elements of the density matrix or the coherences can be
kept vanishingly small to describe the steady state approximately as a thermal equilibrium state. 
The corresponding effective temperature can be determined by the modified
detailed balance condition to reach such a quasi equilibrium state in Eq.~(\ref{Eq.Mot2}) which gives
\begin{eqnarray}
\frac{K_a}{R_{g_0}+R_{g_c}}=\exp{\left(\frac{-\hbar\Omega}{kT_\phi}\right)}.
\end{eqnarray}
The detailed balance between the thermal reservoir at $T_h$ and the photon gas in the resonator is broken but there is
a modified detailed balance between the coherent atomic ensemble and the resonator photons. Accordingly the resonator
can reach a thermal equilibrium at a different temperature $T_\phi$ than $T_h$.

The steady state of the Eq.~(\ref{RateEq.}) yields the average photon number as
\begin{equation}
\bar{n}_{\phi}=\frac{\bar{n}}{1+\bar{n}\frac{R_{g_c}}{ K_a\rho_{aa}}},\label{nbarf}
\end{equation}
where $\bar{n}=(R_{g_0}/K_a\rho_{aa}-1)^{-1}$ is the average photon number in the absence of coherence. Using $ \bar{n}_{\phi}=(\exp(\hbar\Omega/kT_{\phi})-1)^{-1}$, we determine the effective cavity temperature as
\begin{equation} \label{Teff}
T_{\phi}=\frac{T_h}{1+\bar{n}\frac{R_{g_c}}{ K_a\rho_{aa}}},
\end{equation}
by using high temperature approximations $\bar{n}_{\phi}\approx kT_{\phi}/\hbar\Omega$ and $\bar{n}\approx kT_h/\hbar\Omega$ in Eq.~(\ref{nbarf}).

Therefore, the efficiency of the photonic Carnot engine in the case of NLAP becomes
\begin{equation} \label{Eff}
\eta_{\phi}=\eta_{c}-\frac{T_c}{T_h}\bar{n}\frac{R_{gc}}{K_a\rho_{aa}},
\end{equation}
where $\eta_{c}=1-T_c/T_h$ is the Carnot efficiency. Note that for $T_c=T_h$, $\eta_{c}=0$ but $\eta_{\phi}$ could have a positive value for particular values of control parameters $\phi_1,\phi_2,..,\phi_S$.
In order to get further analytical results we will make some simplifying assumptions.

We focus on degenerate NLAP case to proceed analytically, for which 
$E_a=\Omega$, $ E_{b_i}=0, i=1..N$, $\omega_{ab_i}=\Omega$, $\Delta_i=0$, $\omega_{b_ib_j}=0$ and $K_a=2N/\gamma^2$. In addition,
we consider phase locked equal amplitude coherences with $\phi_{ij}=\phi$ and $|\rho_{b_ib_j}|=\lambda$. Hence the coefficients in
Eq.~(\ref{Eq.Mot2}) become $K_{ij}^{\phi_{ij}}=4\cos\phi/\gamma^2$, $R_{g_0}=2NP_g/\gamma^2$ and $R_{g_c}=2N(N-1)\cos\phi \lambda/\gamma\bar{\gamma}$, and hence Eq.~(\ref{RateEq.}) reduces to 
\begin{eqnarray}
\dot{\bar{n}}_{\phi}&=2\mu N[(P_e-P_g+N\xi\lambda)\bar{n}_{\phi}+P_e]-\kappa\bar{n}_{\phi}, \label{DegRateEq.}
\end{eqnarray}
for $N\gg 1$, $\phi=\pi$, where $\mu=rg^2/\gamma^2$, $P_e=\rho_{aa}=\exp(-\beta E_a)/Z$, $P_g=\rho_{b_ib_i}=1/Z$ with $Z=\exp(-\beta E_a)+N$. Here we introduced $\kappa$ and $\xi$, with $|\xi|<1$, as the decoherence rate due to the dissipation in the cavity and a phenomenological decoherence factor due to atomic dephasing, respectively~\cite{quan_quantum-classical_2006}. 
While the dephasing factor is phenomenologically introduced in~\cite{quan_quantum-classical_2006} we provide its rigorous
microscopic derivation in the Appendix.

Steady state solution of Eq.~(\ref{DegRateEq.}) yields an effective temperature given by $T_{\phi}=T_h/(1+F(T_h))$ in the high temperature limit where
\begin{eqnarray}
F(T_h)=\frac{\bar n}{P_e}\left(-N\xi\lambda+\frac{\kappa}{2N\mu}\right),
\end{eqnarray}
with $\bar n=P_e/(P_g-P_e)$.
For small coherence and decoherence terms in $F(T_h)$, an approximate expression can be written for the effective temperature
\begin{eqnarray}
T_{\phi}=T_h\left(1+N^2\xi\lambda\bar n-\frac{\kappa}{2\mu}\bar n\right).
\end{eqnarray}
This result shows that if the reduction of the magnitude of coherence due to dephasing is slower than the quadratic increase with $N$, then the multilevel coherence could be used to beat the decoherence induced by the cavity dissipation. 
The magnitude of coherence $\lambda$ is limited by the positivity requirement 
of the density matrix as well as the thermal equilibrium requirement of the cavity field. The former condition require
$|\rho_{b_ib_j}|\le (\rho_{b_ib_i}\rho_{b_jb_j})^{1/2}$ so that $\lambda \le 1/N$ for $N\gg 1$ 
as $\rho_{b_ib_i}\sim 1/N$ for $N\gg 1$.
Accordingly one can fix the coherence amplitude $\lambda$ as a constant as long as it remains smaller than $1/N$
for the range of $N$ values.
For a realistic number of levels this is not a very restrictive condition. More severe limitation on $\lambda$ is due
to the quasi-equilibrium condition of the photon gas. We will take $\lambda\sim 10^{-6}$ and consider $N\le 40$
in our numerical examinations. In the classical asymptotical limit of $N\rightarrow \infty$ then $\lambda\rightarrow 0$
as $1/N$ and hence
the quadratic scaling reduces to a linear one for which the specific energy becomes a constant as
expected for classical systems. Mesoscopic systems in quantum regime are therefore necessary to exploit the quadratic scaling
in the specific energy.

We note that the coarse grained dynamics is designed on purpose to determine the steady state by rapid convergence using
a numerically efficient dynamical equations. Analytical solution of the mean photon number dynamics for the degenerate case
and with $\lambda=0$, $\kappa=0$, $\xi=1$ gives 
\begin{eqnarray}
\bar n_\phi = \bar n - (\bar n-\bar n_0)\mathrm{e}^{-t/t_{\mathrm{th}}},
\end{eqnarray}
where $t_{\mathrm{th}}=1/2\mu N (P_g-P_e)$ is the thermalization time. We will first discuss the predicted analytical state states by the coarse 
grained master equation in modern resonator settings then examine the exact numerical desription of the dynamics
of the system to verify the analytical results in the subsequent subsections.


\begin{figure}[htb!]
\centering
\includegraphics[width=3.4in]{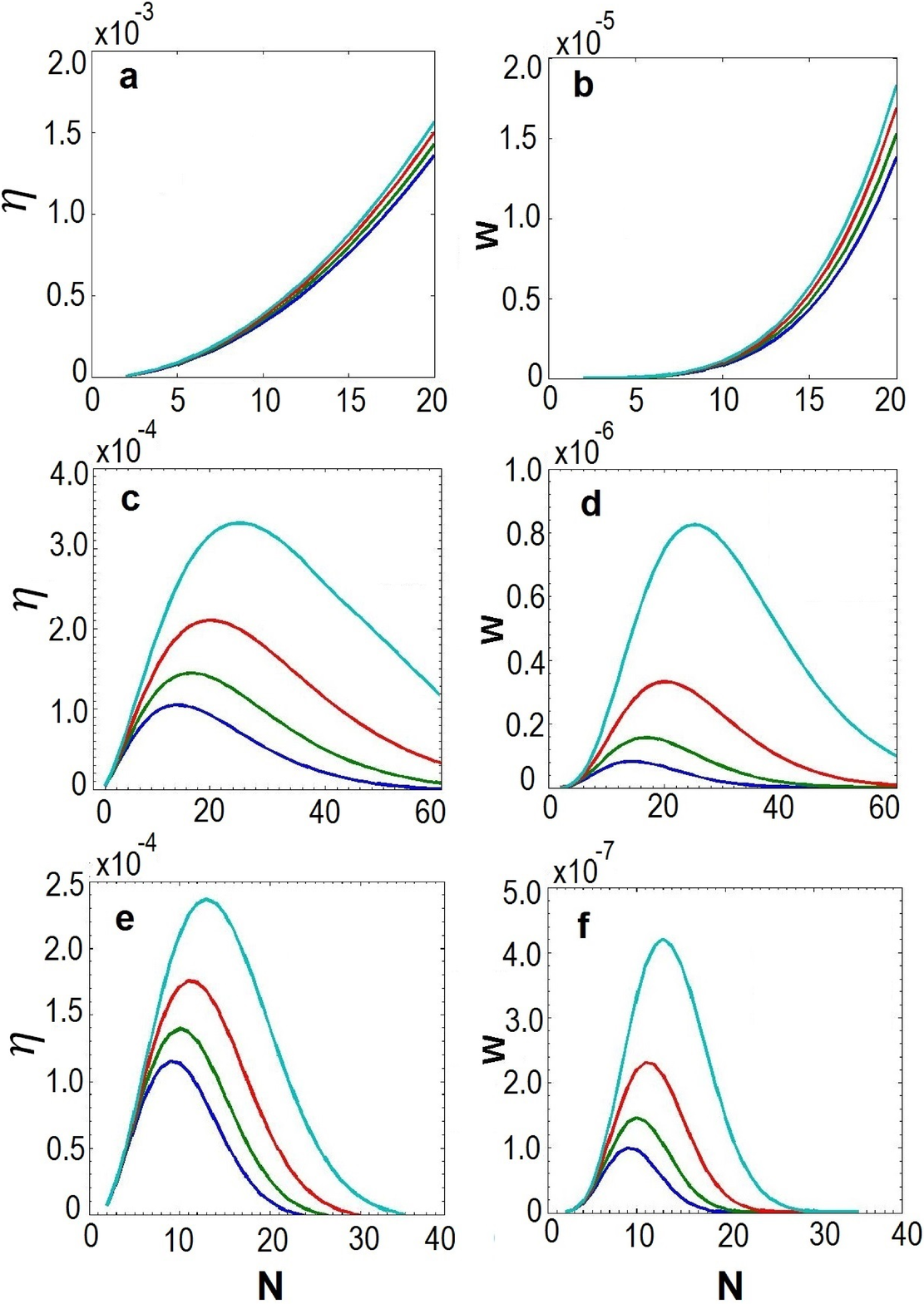}
\caption{\label{fig:fig3}(Color online) Extracted work ($W$) and efficiency ($\eta$) of photonic Carnot engine, with $N+1$ level atom phaseonium (NLAP) fuel,  depending on the number of degenerate coherent ground state levels $N$, for different decoherence factor models (a)-(b) $\xi=\exp{(-x)}$, (c)-(d) $\xi=\exp{(-Nx)}$. (e)-(f) $\xi=\exp{(-N^2x)}$, where $x=\gamma_{\phi}/\gamma$ . Coherence parameter is $\lambda=10^{-6}$ and the initial thermal coherent atomic temperature is $T_h=4$ in units of $\hbar\Omega/k_B$. $x$ values are 0.15, 0.1, 0.05, 0.001 for (a)-(b), 0.14, 0.12, 0.1, 0.08 for (c)-(d) and 0.012, 0.01, 0.008, 0.006 for (e)-(f) in decreasing order for the lower to upper curves respectively. The plots are given for the circuit QED parameters in ~\cite{quan_quantum-classical_2006}. The quantities $g=0.01$, $r=1\times 10^{-4}$, $\kappa=6.25\times 10^{-4}$, and $\gamma=5\times 10^{-6}$, which are the coupling coefficient to the cavity field, atomic injection rate, cavity loss term, and atomic decay respectively, are dimensionless and scaled with  the resonance frequency 
$\Omega\sim 10$ GHz. $\eta$ is dimensionless, $W$ is dimensionless and scaled with $\Omega$.} 

\end{figure} 

\subsection{Analytical results for \\ modern resonator systems}
In the high temperature limit ($T\gg\Omega)$, the entropy change in the isothermal expansion stage is $\Delta S=k\Delta\Omega/\Omega$ and the heat input becomes $Q_{\mathrm{in}}=T_h\Delta S$. The work output at $T_h=T_c$ is found to be $W=Q_{\mathrm{in}}\eta$ where $\eta=\bar n(N^2\xi\lambda-\kappa/2\mu)$, respectively. In superconducting circuit, microwave and optical resonators, it is estimated that $\kappa/2\mu\xi\lambda\sim 10$~\cite{quan_quantum-classical_2006}. $N^2$ should be replaced by $N(N-1)/2$ for smaller number of levels. Accordingly, by using five or more level quantum phaseonium fuel, the working fluid can beat quantum decoherence to harvest positive work.

In Fig.~\ref{fig:fig3}, we plot the work output and efficiency of the photonic Carnot engine with degenerate NLAP fuel, depending on the number of quantum coherent levels. We consider $N$ independent as well as $N$ dependent scaling models~\cite{Yavuz_decohere_2014} for the decoherence factor and take $\xi=\exp{(-x)}$ in Fig.~\ref{fig:fig3}(a)-(b), $\xi=\exp{(-Nx)}$ in Fig.~\ref{fig:fig3}(c)-(d), and $\xi=\exp{(-N^2x)}$ in Fig.~\ref{fig:fig3}(e)-(f), where $x=\gamma_{\phi}/\gamma$ as shown in the Appendix. The plots are given for the circuit QED parameters in ~\cite{quan_quantum-classical_2006}. We consider larger atomic dephasing rates than the typical values to demonstrate its limiting effect on $W$ and $\eta$. The plots indicate that even when there is large dephasing, which can increase with $N$ linearly or quadratically, $W$ and $\eta$ can retain their quadratic power law with $N$ up to a critical $N$. 


\begin{figure}[htb!]
 \centering
\includegraphics[width=3.4in]{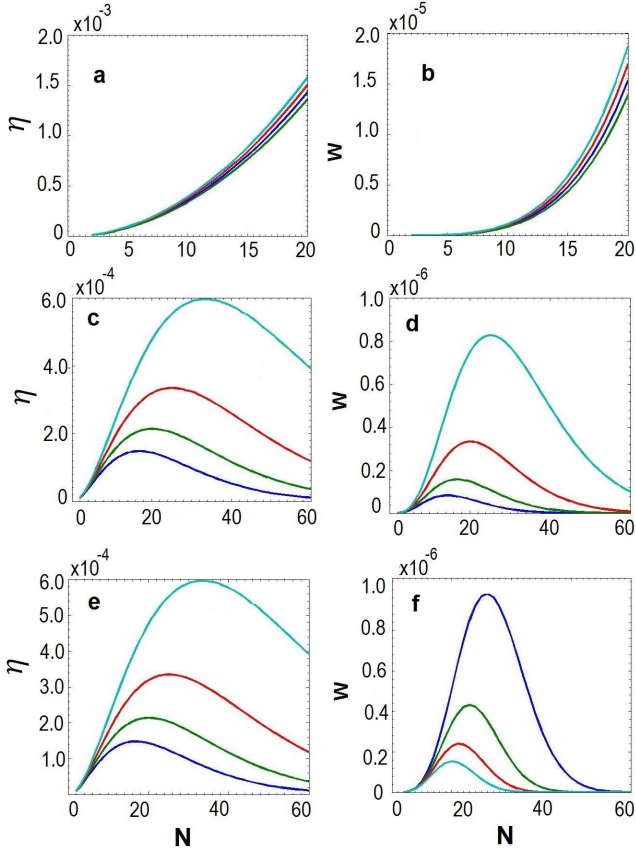}
\caption{\label{fig:fig4}(Color online) Extracted work ($W$) and efficiency ($\eta$) of microwave resonator photonic Carnot engine, with $N+1$ level atom phaseonium (NLAP) fuel,  depending on the number of degenerate coherent ground state levels $N$ for different decoherence factors (a)-(b) $\xi=\exp(-x)$, (c)-(d) $\xi=\exp(-Nx)$, (e)-(f) $\xi=\exp{(-N^2x)}$, where  $x=\gamma_{\phi}/\gamma$. Coherence parameter is $\lambda=10^{-6}$ and the initial thermal coherent atomic temperature is $T_h=4$ in units of $\hbar\Omega/k_B$. $x$ values are 0.1, 0.15, 0.05, 0.001 for (a)-(b), 0.12, 0.1, 0.08, 0.06 for (c), 0.14, 0.12, 0.1, 0.08 for (d), 0.12, 0.1, 0.08, 0.06 for (e) and 0.01, 0.008, 0.006, 0.004 in decreasing order for the lower to upper curves respectively. The parameters $g=9.21\times 10^{-7}$, $r=6.47\times 10^{-5}$, $\kappa=1.96\times 10^{-8}$ and $\gamma=9.54\times 10^{-10}$ are the atom-field coupling coefficient, atomic injection rate, cavity loss term and atomic decay respectively. They are dimensionless and scaled with the typical resonance frequency is $\Omega=51$ GHz \cite{Brune_RydbergPRA_2014}. $\eta$ is dimensionless, $W$ is dimensionless and scaled with $\Omega$.}
\end{figure}

\begin{figure}[htb!]
 \centering
\includegraphics[width=3.6in]{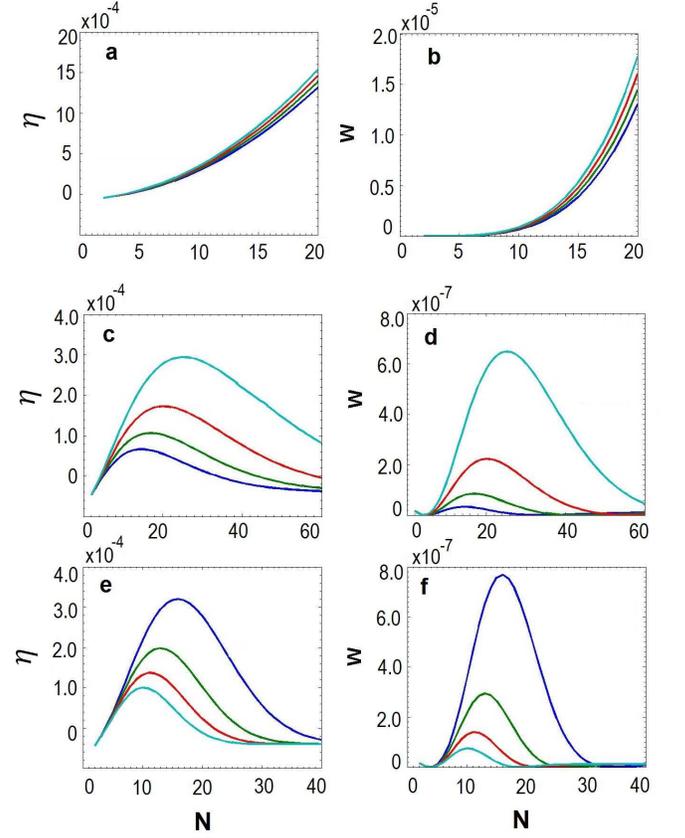}
\caption{\label{fig:fig5}(Color online) Extracted work ($W$) and efficiency ($\eta$) of optical resonator photonic Carnot engine, with $N+1$ level atom phaseonium (NLAP) fuel,  depending on the number of degenerate coherent ground state levels $N$ for different decoherence factors (a)-(b) $\xi=\exp(-x)$, (c)-(d) $\xi=\exp(-Nx)$, (e)-(f) $\xi=\exp(-N^2x)$, where $x=\gamma_{\phi}/\gamma$. Coherence parameter is $\lambda=10^{-6}$ and the initial thermal coherent atomic temperature is $T_h=4$ in units of $\hbar\Omega/k_B$.  $x$ values are 0.15, 0.1, 0.05, 0.001 for (a)-(b), 0.14, 0.12, 0.1, 0.08 for (c)-(d) and 0.01, 0.008, 0.006, 0.004 for (e)-(f) in decreasing order for the lower to upper curves respectively. The parameters  $g=6.28\times 10^{-7}$, $r=8\times 10^{-5}$, $\kappa=2.86\times 10^{-7}$ and $\gamma=4.68\times 10^{-8}$ are the coupling frequency to the cavity field, atomic injection rate, cavity loss term and atomic decay respectively. They are dimensionless and scaled with the typical resonance frequency is $\Omega= 350$ THz \cite{Wolf_Optcoherence_book_1995}. $\eta$ is dimensionless, $W$ is dimensionless and scaled with $\Omega$.} 
\end{figure}

Similar results are found for the cases of optical and microwave cavities. We see both in Fig.~\ref{fig:fig4}(a)-(b) and Fig.~\ref{fig:fig5}(a)-(b) that when dephasing is independent of $N$, the work output and efficiency increases quadratically with the number of coherent levels. If the dephasing rate is increasing linearly with $N$ as in Fig.~\ref{fig:fig4}(c)-(d) and Fig.~\ref{fig:fig5}(c)-(d), or if it is increasing quadratically with $N$  as in Fig.~\ref{fig:fig4}(e)-(f) and Fig.~\ref{fig:fig5}(e)-(f), the work output and efficiency of the photonic engine is enhanced quadratically with the number of coherent levels only up to critical $N$. beyond which the work output and efficiency decays exponentially due to the dominating effect of decoherence. 


\subsection{Numerical verification of the theory}

In order to perform a faithful numerical simulation of a typical set up described in the theory, we investigate the injection process in detail. We assume a regular atomic injection of Rydberg atoms into a Fabry-Perot cavity~\cite{Manupilating_Raimond_2001} with an atomic interaction time $\tau$ with the cavity field and an empty cavity time $\tau_0$ such that $1/r=\tau + \tau_0$ where $r$ is the injection rate. During the time interval $\tau$, the  hamiltonian is 
\begin{eqnarray}\label{transient Hamilt.}
H=\omega_a\ket{a}\bra{a}+\Omega\hat{a}^{\dagger}\hat{a}+g(\sum_{i=1}^N \ket{a}\bra{b_i}\hat{a}+H.c.)
\end{eqnarray}
while for the time interval $\tau_0$, $H=\Omega\hat{a}^{\dagger}\hat{a}$ ($\hbar=1$ and $\omega_{b_i}=0$ 
for degenerate ground state levels).

We choose injection time $1/r=1/(N_{ex}\kappa)$ where $\kappa$ is the cavity decay rate and $N_{ex}$ is the number of atoms kicking the cavity field in the photon lifetime. The time elapsed when cavity is empty $\tau_0$ can be related to the interaction time such that $\tau_0=N_{em}\tau$. Thus, we can write $1/r=\tau(1+N_{em})$. Here, $N_{em}$ is a factor introduced to measure $\tau_0$ in terms of $\tau$ so that $N_{em}=1/(N_{ex}\kappa\tau)-1$. 

We solve the master equation by numerical methods and compare the results with the developed theory. We use QuTip package \cite{Qutip_Nori_2012} in Python software to solve the master equation. We perform single atomic injection in two steps. First step is the atom-cavity field interaction (during $\tau$) and the second one is the free cavity field  evolution (during $\tau_0$). The master equation for the first step is written under Markov and Born-Markov approximations as \cite{Zoller_Q.Noise_2004}
\begin{eqnarray}\label{LindbladEq}
\dot{\rho}=-i[H,\rho]+\gamma\sum_m^{N+1} \mathcal{L}[L_m^{\gamma}]+\frac{\gamma_{\phi}}{2}\sum_n^N \mathcal{L}[L_n^{\phi}]
\end{eqnarray}
where last two terms stand for pure spontaneous emission and pure dephasing  \cite{Synthetic_Girvin_2011,OptimalControl_NLevel_Pötz_2005,Constraint_Relax_Solomon_2004,Pure_dephasing_Hakonen_2012}, respectively. Here, $\mathcal{L}[x]=(2x\rho x^{\dagger}-xx^{\dagger}\rho-\rho x^{\dagger}x)/2$ is a Liouvillian superoperator in Lindblad form and $L_m^{\gamma}=\ket{r}\bra{\alpha_m}$, $L_n^{\phi}=\ket{b_n}\bra{b_n}$. We include an auxiliary state $\ket{r}$ to the atomic state space to model the decay of the excited and the lower levels. Presence of $\ket{r}$ is not 
altering the initial phaseonium state. The auxiliary state is 
unpopulated and at the lower level energy. Its use allows for faithful simulation of the excited state and the degenerate ground state $(\alpha_m=a,b_1,..,b_N$) decay equations in Eq.~(\ref{Eq15}). This decay model is already used in the original master equation developed for the two-level phaseonium engine~\cite{Rostovtsev_extracting_2003}. 
Different decay models, for example decay of excited level to the lower levels are employed for other systems such as many atom superradiant Otto engine~\cite{Superradiant_Otto_2015} and similar effect of beating decoherence with scaling up coherence is found. Present contribution discusses the original photo-Carnot engine~\cite{scully_extracting_2003} as well as the objection to its feasibility 
due to dephasing, phenomenologically described by factor $\xi$~\cite{quan_quantum-classical_2006}. Our introduction of $\gamma$ is an additional decoherence channel not included in Ref.~\cite{quan_quantum-classical_2006}.  We have found that $\gamma_\phi$ can be analytically expressed in terms of $\xi$ (see Appendix for details). The microscopical master equation approach describes 
both the
original photo-Carnot engine coarse-grained master equations with~\cite{quan_quantum-classical_2006} and 
without~\cite{Rostovtsev_extracting_2003} dephasing and generalizes them to the multilevel case; as our mesoscopic 
master equation, Eq.~\ref{Eq.Mot2}, does analytically. 

In the second step, cavity decay $(\kappa)$ is present during the time interval $\tau_0$ in accordance with the key assumptions of micromaser theory~\cite{filipowicz_theory_1986} and the corresponding  master equation is
\begin{eqnarray}\label{LindbladEq}
\dot{\rho}=-i[H,\rho_f]+\kappa\mathcal{L}[\hat{a}].
\end{eqnarray}


\begin{figure}[htb!]
 \centering
\includegraphics[width=3.2in]{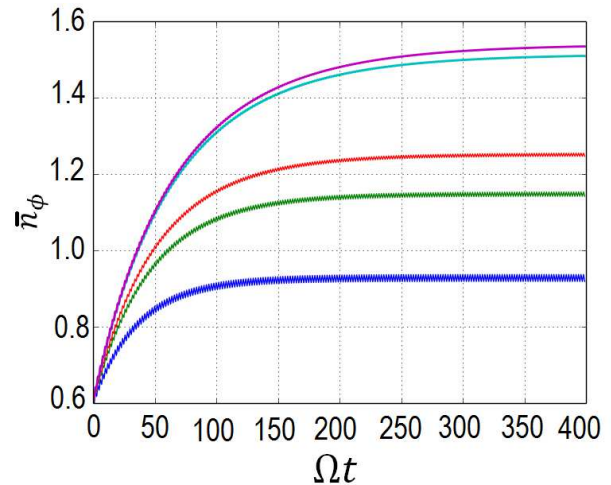}
\caption{\label{fig:fig6}(Color online) Time evolution of average photon number during thermalisation process of PCE field while regular injection of 2LAP depending on different $N_{ex}$ parameters. $N_{ex}$ values are 4500, 1500, 150, 100 and 50 in decreasing order for the upper to lower curves respectively. Coherence parameter is $\lambda=10^{-3}$, initial field temperature $T_f=1$ and temperature of thermal coherent atoms is $T_h=2$ in units of $\hbar\Omega/k_B$. The resonant field frequency is $\Omega=51$ GHz, cavity quality factor is $Q=2\times 10^{10}$, atom cavity field interaction time is $ \tau=10$ $\mu$s, atom decay rate is $\gamma=33.3 $ Hz, atom dephasing rate is $\gamma_{\phi}=3.3 $ Hz and atom cavity field coupling is $g=50 $ kHz. Time is dimensionless and scaled with $\Omega$.} 
\end{figure}  
 In Fig. \ref{fig:fig6}, we present the thermalisation process of the  cavity field depending on different $N_{ex}$ values by depicting the photon number versus scaled time. Physical parameters~\cite{quan_quantum-classical_2006} are given in the figure caption consistent with the Rydberg atoms in a superconducting Febry-Perot cavity~\cite{Manupilating_Raimond_2001}. For low values of $N_{ex}$ which corresponds to large $N_{em}$, we have zigzag like curves and for high values of $N_{ex}$, we have smoother lines. Average photon number $\bar{n}_{\phi}$ converges to the theoretical value for $N_{em}=12\times 10^3$. Thermalisation time is much longer than the convergence rate of coarse grained master equation and the microscopic exact method is much more costly numerically. 


\begin{figure}[htb!]
 \centering
\includegraphics[width=3.2in]{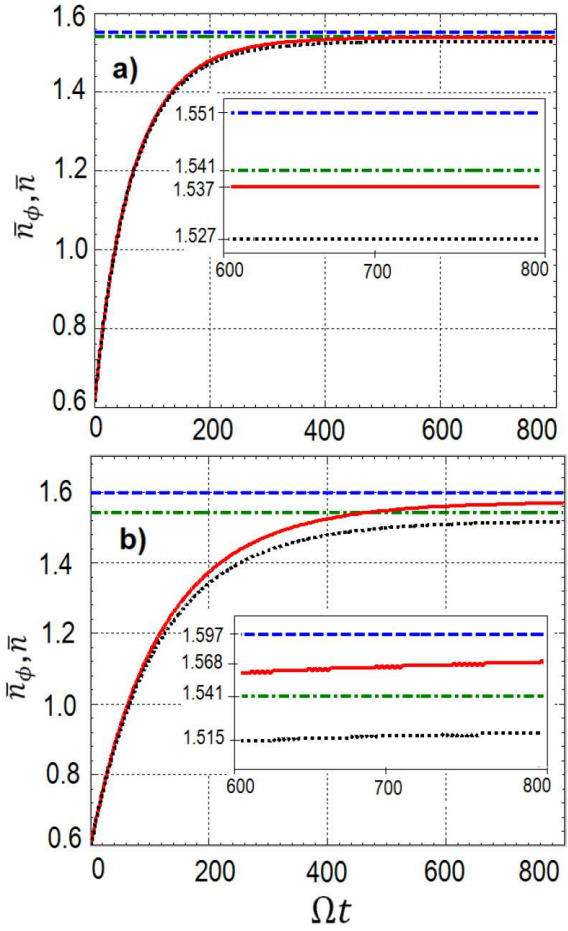}
\caption{\label{fig:fig7}(Color online) Comparison of time evolution of average photon number in presence of thermal and coherent atom injection with $N=2$ and $N=4$. The horizontal dashed and dashed-dotted lines stands for the analytical $\bar{n}_{\phi}$ and $\bar{n}$ values in absence of loss mechanisms. Solid and dotted lines stands for time evolution of $\bar{n}_{\phi}$ and $\bar{n}$ respectively in presence of dissipation channels. $N_{ex}=4000$ for both subplots corresponding to $\tau_0=90$ $\mu$s. Insets magnifies the lines between $\Omega t=600$ and $\Omega t=800$. All the remaining parameters are the same with that of Fig.~\ref{fig:fig6}. Time is dimensionless and scaled with $\Omega$.} 
\end{figure}


In  Fig.~\ref{fig:fig7}(a)-(b), we express the effect of the number of degenerate atomic ground state levels $N$ of the coherent atoms against the dephasing and the cavity loss mechanisms. Horizontal dotted and dotted-dashed lines  are the  analitical values of average photon numbers ( $\bar{n}_{\phi}$, $\bar{n}$) of the no loss case for each $N$. 
In  Fig.~\ref{fig:fig7}(a) when decoherence channels are open, average photon number saturates below the analytical values for $N=2$ in accordance with the argument that 2LAP phaseonium can not beat decoherence ~\cite{quan_quantum-classical_2006}. Average photon number exceeds $\bar{n}$ for $N=4$ in Fig.~\ref{fig:fig7}(b) by keeping all the other parameters same. Thus, we show that the decoherence can be beaten by using higher $N$. To make the effect more visible in the figures, we take larger coherence magnitude, 
$\lambda=10^{-3}$.

\begin{figure}[htb!]
 \centering
\includegraphics[width=3.5in]{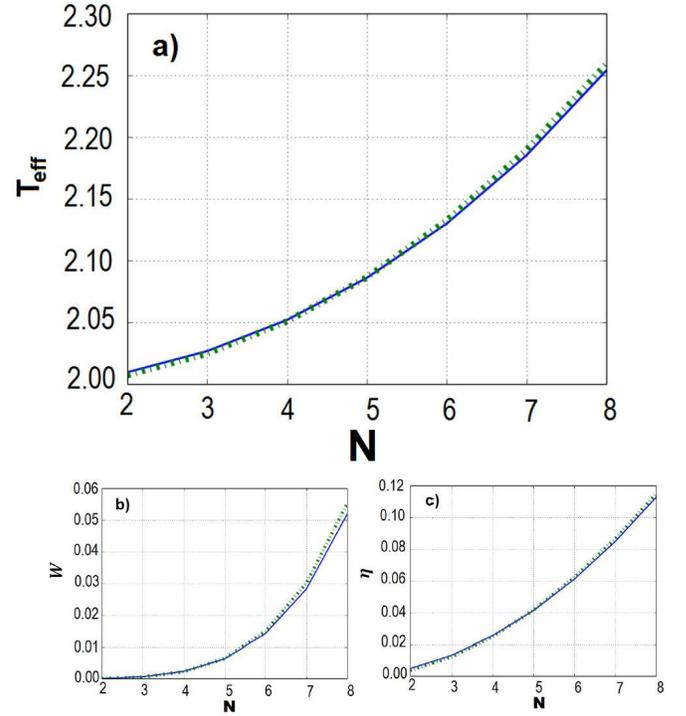}
\caption{\label{fig:fig8}(Color online) Comparison of numerical results (dashed line) with developed theory (solid line) of effective field temperature 
 T$_{\mbox{\scriptsize eff}}$, harvested work $W$ and efficiency $\eta$ respectively depending on the number of degenerate ground state levels $N$. $N_{ex}=12\times 10^3$ and corresponding $\tau_0=2.01$ $\mu$s. All the remaining parameters are the same that of Fig.~\ref{fig:fig6}. T$_{\mbox{\scriptsize eff}}$, $W$ and $\eta$ is dimensionless and scaled with $\Omega$.} 
\end{figure}

We also compare the consistency of effective field temperature T$_{\mbox{\scriptsize eff}}$, harvested work $(W)$ and efficiency $(\eta)$  versus $N$ in Fig.~\ref{fig:fig8} (a), (b), (c)  between developed analytical  and numerical results when dissipation channels open. We observe a good consistency between numerical and theoretical results in steady state. 

\subsection{Preparation of the NLAP \\and its energy cost}\label{Cost} 

Typical methods to generate quantum superposition states, such as pulse area, adiabatic passage, or STIRAP techniques~\cite{amniat-talab_superposition_2011,bevilacqua_implementation_2013}, 
utilize optical pulses interacting with the atomic system to transfer an initial atomic state to a target one. The initial and target
quantum states are known and hence one can easily determine the required unitary transformation between them. 
Physical implementation of the required propagator is however a much more
challenging problem than the calculation of the transformation matrix. An efficient strategy to sythesize the transformation matrix is to decompose it into a product of matrices, representing interacting steps that 
can be implemented by using optical pulses coupled to the atom. 

An arbitrary $N$ dimensional unitary matrix ${\bf U}(N)$ can be decomposed into so called $N$ generalized quantum Householder reflection (QHR) matrices or $N-1$ standard QHRs and a phase gate~ \cite{ivanov_engineering_2006,ivanov_navigation_2007}. 
A generalized QHR is defined by
\begin{eqnarray}\label{GenReflct}
{\bf M}(\nu;\phi)={\bf I}+(e^{i\phi}-1)\ket{\nu}\bra{\nu}
\end{eqnarray} 
where ${\bf I}$ is the identity operator and the $\ket{\nu}$ is the normalized column vector with dimension $N$, same with the number of the pulses, and $\phi$ is an arbitrary phase factor. The decomposition of ${\bf U}(N)$ in terms of generalized QHRs can be written as
\begin{eqnarray}\label{Decomp}
{\bf U}={\bf M}(\nu_1;\phi_1){\bf M}(\nu_2;\phi_2)...{\bf M}(\nu_N;\phi_N).
\end{eqnarray}
For $\phi=\pi$, Eq.~(\ref{GenReflct}) reduces to ${\bf M}={\bf I}-2\ket{\nu}\bra{\nu}$ which is the standard QHR.
The interaction represented by each Householder matrix can be described by a propagator which can be determined by the
Morris-Shore transformation~\cite{saadati-niari_creation_2014}. 

Our N+1 level atom coupled to N optical pulses in a fan shaped transition scheme, or so called N-pod model, 
is a generic model that is used to discuss generation of arbitrary multilevel superposition states. Under the Morris-Shore transformation, 
the lower levels of the atom are grouped into a single bright level coupled to an effective single pulse and N-1 dark levels uncoupled from the optical pulses. The propagator is then easily determined in this Morris-Shore basis. Back transformation to the original basis gives the full propagator, or the generalized QHR matrix. Both the number of QHR steps and the number of pulses used in each step are in the order of 
$N$, therefore the total number of pulses to be used to generate the target state would be in the order of $N^2$. This shows that the energetic cost of preparation of the target
state scales with $N^2$, same with the work and efficiency scaling in the corresponding photonic Carnot engine. 

The preceding discussion is applicable to the case of mixed states as well, for which 
the normalized vectors of generalized QHRs are defined as \cite{ivanov_engineering_2006}
\begin{eqnarray}\label{NormalisedVect}
\ket{\nu_i}=\frac{1}{e^{-i\phi_i}-1}\sqrt{\frac{2\sin{(\phi/2)}}{|1-u_{ii}|}}(\ket{u_i}-\ket{e_i}).
\end{eqnarray}
Here, $u_i$ is the $i^{th}$ column of ${\bf U}(N)$, $\ket{e_i}=[0,..,1^{ith},..0]^T$ and $\phi$ is an arbitrary phase where 
$\phi_i=2\mathrm{arg}(1-u_{ii})-\pi$. It's shown that for an N-Pod system any standard QHR ${\bf M}(\nu)$ 
can be realized by single pulses with an rms pulse area $A=2\pi$ \cite{ivanov_navigation_2007}. The corresponding unitary transformation
can only link the mixed states with the identical dynamical invariants. In our case we
consider initial thermal states out of a hohlarum transformed to a final state with small coherences. The initial and final states
would then possess different spectral decompositions so that they cannot be unitarily connected. A resolution to this
is suggested to exploit decoherence channels such as spontaneous emission or pure dephasing, in combination with the 
unitary transformation~\cite{ivanov_engineering_2006,ivanov_navigation_2007}. We will not follow this route but use
an alternative, which allows for a fully unitary procedure to generate desired coherences. As the coherences contribute
additively, to exploit their scaling advantage we do not need an exact state but an approximate one would be sufficient.
Accordingly, we can simply consider an approximate approach and do not specify
an exact value for the coherences. We only need to keep them small enough 
to ensure slightly out of thermal equilibrium final state. We
illustrate our strategy for $N=2$ case, and suggest that in principle larger NLAP can be generated by 
straightforward extension
of this technique. Unitarity of our procedure also makes the details of generation process immaterial for the cost estimation.
The cost would be the same for other unitary equivalent processes to generate same states. 

Initial state of the atom out of the hohlarum at $T_h$ is the thermal density matrix 
\begin{eqnarray}\label{ThermalDm}
\rho_{\mathrm{th}}=\frac{1}{Z}e^{-\beta H}=\sum_{n=1}^{N+1}P_n\ket{\Psi_n}\bra{\Psi_n}
\end{eqnarray}
where $\beta=1/k_BT_h$ inverse temperature $(k_B=1)$ and $Z=\mathrm{Tr} \mathrm{e}^{-\beta H}$ is the partition function, with $H=\sum_{i=1}^{N+1}\hbar\omega_i\ket{i}\bra{i}$ being the atomic Hamiltonian. Taking $N=2$  and $T_h=2$, we find 
$\rho_{\mathrm{th}}=\mathrm{diag}(0.327, 0.384, 0.384)$. Target density matrix $\rho_c$ is taken to be  
\begin{eqnarray}\label{CoherentDm}
\rho_c=\left(\begin{array}{ccc} 0.327 & 0 & 0\\0 & 0.384 & 0.000001\\0 & 0.000001 & 0.384\end{array}\right),
\end{eqnarray}
where the off-diagonal elements between degenerate ground state levels are taken real and much smaller than the 
diagonal elements that are equal to those of the $\rho_{\mathrm{th}}$. 
The initial and final density matrices have distinct dynamical invariants in their 
spectral decompositions and hence they cannot be
linked by a coherent evolution. Let us assume however an approximate link such that 
$\rho_c\approx {\bf U}\rho_{\mathrm{th}}{\bf U}^{\dagger}$. The unitary transformation ${\bf U}$ can be determined
from the matrix that diagonalize $\rho_c$ and found to be  
\begin{eqnarray}\label{TildeU}
{\bf U}=\left(\begin{array}{ccc} 1 & 0 & 0\\0 & -0.707 & 0.707\\0 & 0.707 & 0.707\end{array}\right).
\end{eqnarray}
Writing $\tilde{\rho_c}={\bf U}\rho_{ \mathrm{th}}{\bf U}^{\dagger}$, 
the fidelity between $\rho_c$ and $\tilde{\rho_c}$ is determined by ${\cal F}(\rho_c,\tilde{\rho_c})=|Tr\sqrt{\sqrt{\rho_c}\tilde{\rho_c}\sqrt{\rho_c}}|^2$ and found be ${\cal F}(\rho_c,\tilde{\rho_c})\simeq 1$. Thus, $\tilde{{\bf U}}$ can be approximately be used as the unitary operator linking the initial and the target density matrices. 

$\tilde{\bf U}$ can be synthesized by using two standard QHRs and a phase gate as
\begin{eqnarray}\label{DecomposeTildeU}
\tilde{{\bf U}}={\bf M}(\nu_1;\phi_1){\bf M}(\nu_2;\phi_2){\bf M}(\nu_3;\phi_3)
\end{eqnarray}
where$\phi_1=\phi_2=\phi_3=\pi$ and ${\bf M}(\nu_3;\phi_3)=\Phi(0,0,\Phi_3)$ is a one-dimensional phase gate. By using  Eq.~(\ref{NormalisedVect}) one finds the normalized column vectors to be $\ket{\nu_1}=[0, 0, 0]^T$, $\ket{\nu_2}=[0, 0.924, -0.383]^T$ . The example for $N=2$ here illustrates the basic principles to generate arbitrary NLAP. One can use $N$ pulses for each
$N-1$ standard QHRs and a phase gate to synthesize a unitary transformation matrix, which is the diagonalization matrix of 
a quasi equilibrium thermal state at $T_h$ with small coherences. 
When the pulses applied to the actual atom out of hohlarum at $T_h$, the final
state will be approximately the same with the target state used to determine the properties of the pulses.

We can now estimate the energy cost $U_c$ of generating $\rho_c$ using $U_c=N^2U_p$ 
where $N^2$ is the total number of pulses used in the
QHR technique and $U_p$ is the energy of a single pulse. $U_p$ can be determined from the pulse area $A$ as in ~\cite{Superradiant_Otto_2015}. For a square pulse of
duration $\tau_p$ and amplitude $E_p$ we have $A=dE_p\tau/\hbar$, where $d$ is the magnitude of the dipole moment 
\begin{eqnarray}\label{eq:dipoleMoment}
d=\sqrt{\frac{3\pi\epsilon_0\hbar c^3\gamma}{\Omega^3}},
\end{eqnarray}
where $\epsilon$ is the vacuum permittivity and $c$ is the speed of light. Taking the pulse area $A=2\pi$ we find
\begin{eqnarray}\label{eq:pulseAmp}
E_p=\frac{2\pi\hbar}{\tau_p}\sqrt{\frac{\Omega^3}{3\pi\epsilon_0\hbar c^3\gamma}}.
\end{eqnarray}
The intensity of the pulse is given by $I_p=c\epsilon_0|E_p|^2/2$. The pulse energy in a beam of radius $r_b$ can be estimated
by $U_p = \pi r_b^2 I_p\tau_p$. Using $\Omega=2\pi c/\lambda$ and $\zeta=\lambda/2\pi r_b$, where $\lambda$ and $\zeta$ are the 
wavelength of the optical field and the radial beam divergence, respectively, we find 
\begin{eqnarray}
U_p = \hbar \Omega \frac{\pi^2}{6}\frac{1}{\tau_p\gamma}\frac{1}{\zeta^2}.
\end{eqnarray}
Taking $1/\tau_p\gamma\sim 2$ and $\zeta\sim 0.5$~\cite{Superradiant_Otto_2015}, we find $U_p\sim 12\hbar\Omega$. 

Total energy cost to reach steady state per cycle is $U_{\mathrm{ss}}=mU_c=mN^2U_p$ where $m=r\Delta t_s$. Here, $m$ is the number of atoms needed for thermalisation, $r$ is the injection rate, and $\Delta t_s$ is the time elapsed to reach the steady state. In our results
we have found that the harvested work per cycle is much less than the resonance energy, $W\ll \hbar\Omega$, thus 
the generation cost of NLAP fuel is several orders of magnitude larger than the harvested work 
$U_{\mathrm{ss}}\gg W$, which confirms that the second law of thermodynamics obeyed in photonic Carnot engine with NLAP. 
Generation cost is not included in the thermodynamic efficiency but it can be significant figure of merit in the round trip efficiency.
To make such photonic Carnot engines more appealing for certain applications, it is necessary to increase their round trip
efficiency as well. For that aim one could consider the cases of larger compression ratios ($\Delta\Omega\gg\Omega$) and 
high operation temperatures ($k_BT_h\gg \hbar\Omega$). Our focus here is on the discussion if such engines
can operate under decoherence. Despite the negative conclusions for two level phaseonium case~\cite{quan_quantum-classical_2006}, we have
found that larger phaseonium fuel allows for operational photonic Carnot engines. 
The question of how to increase their round trip efficiency
requires further analysis which is beyond the scope of present contribution. 
\section{Conclusions}\label{Conclude}
Summarizing, we examined scaling of work and efficiency of a quantum heat engine with the number of quantum resources. Specifically, we considered a photonic Carnot engine with a multilevel phaseonium quantum fuel. We derived a generalized master equation for the cavity photons, which forms the working fluid of the engine, and determined the steady state 
photon number to calculate the work output and thermodynamic efficiency. We find that they scale quadratically with the number of quantum coherent levels $N$. We examined the case of degenerate levels to get analytical results and to examine scaling laws against decoherence due to cavity dissipation and atomic dephasing. We verified our analytical results with detailed numerical methods and have shown 
consistency of coarse grained analytical results with the microscopical numerical approach. Generation of multilevel phaseonium
fuel using Morris-Shore transformation determined quantum Househoılder reflection technique as well as its cost are examined. 
Using typical parameters in modern resonator systems, such as circuit QED, our calculations reveal that decoherence due to cavity dissipation could be overcome by the multilevel quantum coherence even in the presence of large dephasing rate. If the dephasing rate increases with $N$, then work and efficiency can still overcome the decoherence and retain their $N^2$ scaling up to a critical number of coherent levels.
\section{Acknowledgements}
Authors warmly thank N. Allen, A. Imamoglu and I. Adagideli for illuminating discussions. 
Authors acknowledge support from Ko\c{c} University and Lockheed Martin University Research Agreement.

\appendix
\setcounter{section}{1}
\section*{Appendix} \label{Appdx}

We generalize micromaser mesoscopic master equation treatment \cite{Rostovtsev_extracting_2003} applied for three-level phaseonium engine \cite{scully_extracting_2003} to multi-level case. In $N$LAP model, the Hamiltonian of the whole system is $H=H_0+H_I^k$ where label $k$ implies injected $k^{th}$ arbitrary atom. We adopt the notation of ref. \cite{Bergou_injectedCoherence_1989} for the Hamiltonian and relevant quantities. Here, $H_0=\hbar\omega_a\ket{a^k}\bra{a^k}+\hbar\sum_{i=1}^{N}\omega_{b_i}\ket{b_i^k}\bra{b_i^k}$ and

\begin{eqnarray}
H_I^k=\hbar g\sum_{i=1}^{N} \ket{a^k}\bra{b_i^k}\hat{a}e^{-i\Omega t}+H.c.\label{EQ-int}
\end{eqnarray}
in the interaction picture where $\omega_1$, $\omega_{b_i}$ are atomic energy levels, $g$ is the atom-field coupling coefficient and $\Omega$ is the single mode cavity frequency. Here, we assume all levels coupled to the excited one with the same coefficient $g$. The equation of motion of overall system is 

\begin{equation}
\dot{\rho}=-\frac{i}{\hbar}[H,\rho]+\mathcal{L}_A[\rho]+\mathcal{L}_f[\rho]\label{Overall-EQ-Mot}
\end{equation}
where $\mathcal{L}_A[\rho]$ and $\mathcal{L}_f[\rho]$ are the Liouvillian superoperators expressed in the main text corresponding to atomic and field degrees of freedom respectively.

The equation of motion of the radiation field which is the working substance of the heat engine can be found by tracing out atomic part as
\begin{align}
\dot{\rho}_{nn}=-\frac{i}{\hbar}\sum_{k}(\mbox{Tr\scriptsize at}[H^k,\rho^k]_{nn})\nonumber\\
+\mbox{Tr\scriptsize at}\mathcal{L}_A[\rho]_{nn}+\mbox{Tr\scriptsize at}\mathcal{L}_f[\rho]_{nn}\label{EQ-Mot}
\end{align}
where $\dot{\rho}_{nn}=\bra{n}\dot{\rho}\ket{n}$. Here, $[H,\rho]_{nn}=\bra{n}(H\rho-\rho H)\ket{n}=
\sum_{m}\bra{n}H\ket{m}\bra{m}\rho\ket{n}-\sum_{m}\bra{n}\rho\ket{m}\bra{m}H\ket{n}=\sum_{m}(H_{nm}\rho_{mn}-\rho_{nm}H_{mn})$.
In micromaser theory, due to the short atom-cavity interaction time, last term, cavity decay is usually ignored when the atom is inside the cavity. The second term is treated perturbatively and will be considered for zeroth order in g. Here, we will keep it but assume it can be treated independently. One may write the partial trace operation over atomic degrees of freedom for a random single atom as, 

\begin{align}
\mbox{Tr\scriptsize at}[H,\rho]_{nn}&=\sum_{\alpha}\bra{\alpha,n}[H,\rho]\ket{\alpha,n}\nonumber\\
&=\bra{a n}[H,\rho]\ket{a n}+\sum_{i=1}^{N}\bra{b_i n}[H,\rho]\ket{b_i n}\label{Ptrace}
\end{align}
where $\alpha$  are the atomic basis as expressed at the right hand side. Each term of Eq.~(\ref{Ptrace}) can be calculated by using the selective rules of the Hamiltonian between certain levels $n$ and $m$; for instance inserting for the first term of Eq.~(\ref{Ptrace}) we have, 

\begin{align}
\bra{a n}[H,\rho]\ket{a n}&=\bra{a n}(H\rho-\rho H)\ket{a n}\nonumber\\
&=\sum_{\alpha' m}\{\bra{a n}H\ket{\alpha' m}\bra{\alpha' m}\rho\ket{a n}\nonumber\\
&-\bra{a n}\rho\ket{\alpha' m}\bra{\alpha' m}H\ket{a n}\}\nonumber\\
& =\sum_{\alpha' m}\{H_{an,\alpha'm}\rho_{\alpha'm,an}-\rho_{an,\alpha'm}H_{\alpha'm,an}\}.
\label{Eq4}
\end{align}
$H_I$ is the hamiltonian Eq.~(\ref{EQ-int}) to be inserted into Eq.~(\ref{Eq4}) which can be  written as,
\begin{eqnarray}
H_I=\hbar g\hat{R}_{+}\hat{a}e^{-i\Omega t}+\hbar g\hat{R}_{-}\hat{a}^{\dagger}e^{i\Omega t},\label{Eq5R}
\end{eqnarray}
where $\hat{R}_{+}=\sum_{i=1}^{N}\ket{a}\bra{b_i}$ and $\hat{R}_{-}=\hat{R}_{+}^{\dagger}$. Expressing the terms of Eq.~(\ref{Eq4}) conveniently, we write 
\begin{align}
(\hat{R}_{+})_{a\alpha'}&=\bra{a}(\ket{a}\bra{b_1}+...+\ket{a}\bra{b_N})\ket{\alpha'}\nonumber\\
&=(\delta_{b_1\alpha'}+...+\delta_{b_N\alpha'})\label{Eq6}
\end{align}
and $(\hat{R}_{-})_{a\alpha'}=0$.  
Besides, $\hat{a}_{nm}=\bra{n}\hat{a}\ket{m}=\sqrt{m}\bra{n}\ket{m-1}=\sqrt{m}\delta_{n,m-1}$. Substituting these terms into the first part of Eq.~(\ref{Eq4}) we have $\sum_{\alpha' m}\{H_{an,\alpha'm}\rho_{\alpha'm,an}\}=\hbar g e^{-i\Omega t}\sqrt{n+1}(\rho_{b_1 n+1,a n}+...+\rho_{b_N n+1,a n})$. The second part of Eq.~(\ref{Eq4}) is simply the complex conjugate. Hence the first term of Eq.~(\ref{Ptrace}) is
\begin{align}
\bra{a n}[H,\rho]\ket{a n}&=\hbar g e^{-i\Omega t}\sqrt{n+1}\sum_{i=1}^{N}\rho_{b_i n+1,an}-c.c.\nonumber\\
 \label{Eq7}
\end{align}

The second term of Eq.~(\ref{Ptrace}) would be calculated by similar considerations. We can write $\bra{b_i n}[H,\rho] 
\ket{b_i n}=\sum_{\alpha' m}\{H_{b_i n,\alpha'm}\rho_{\alpha'm,b_i n}\}$. In this case $(\hat{R}_{+})_{b_i\alpha'}=0$, 
$(\hat{R}_{-})_{b_i\alpha'}=\bra{a}\ket{\alpha'}=\delta_{a\alpha'}$ and $\hat{a}_{nm}^{\dagger}=\bra{n}\hat{a}^{\dagger}\ket{m}=\sqrt{m+1}\bra{n}\ket{m+1}=\sqrt{m+1}\delta_{n,m+1}$. Then, $\bra{b_i n}[H,\rho]\ket{b_i n}=\hbar g e^{i\Omega t}\sqrt{n}\rho_{an-1,b_in}-c.c.$ and finally the second term of Eq.~(\ref{Ptrace}) becomes,
\begin{eqnarray}
\sum_{i=1}^{N}\bra{b_in}[H,\rho]\ket{b_in}=\hbar g e^{i\Omega t}\sqrt{n}\sum_{i=1}^{N}\rho_{an-1,b_in}-c.c.\nonumber\\\label{Eq8bi}
\end{eqnarray} 
Inserting these results into Eq.~(\ref{EQ-Mot}) and after some rearrangements we have the field equation of motion, 
\begin{align}
\dot{\rho}_{nn}&=-g\sum_k\{(i\sqrt{n+1}e^{-i\Omega t}\sum_{i=1}^{N}\rho_{b_i n+1,an}^{k}\nonumber\\
&-i\sqrt{n}e^{-i\Omega t}\sum_{i=1}^{N}\rho_{b_i n,an-1}^{k})+c.c.\}\nonumber\\
&+\mathcal{L}_f[\rho_f]_{nn}.\label{9EqMot}
\end{align}
We use $\rho_f\equiv\rho$ hereafter for simplicity. Here,

\begin{eqnarray}
\mathcal{L}_f[\rho]_{nn}&=\bra{n}\frac{\kappa}{2}(2\hat{a}\rho\hat{a}^{\dagger}-\rho\hat{a}^{\dagger}\hat{a}-\hat{a}^{\dagger}\hat{a}\rho)\ket{n}\nonumber\\
&=\kappa\left\lbrace(n+1)\rho_{n+1,n+1}-n\rho_{nn}\right\rbrace.\label{Louvil-Rho}
\end{eqnarray}

In order to proceed the calculation, any single term in the summation of Eq.~(\ref{9EqMot}) should be calculated and inserted therein. The terms can be obtained by the integration of corresponding equation of motions by using the selecetive rules of the Hamiltonian as expressed above. We evaluate atomic equation of motion $0^{th}$ order in $g$ first, we have

\begin{align}
\dot{\rho}_A&=-\frac{i}{\hbar}[H_A,\rho_A]+\gamma_{\alpha,\alpha'}\sum_{\alpha,\alpha'}\mathcal{L}[L_{\alpha,\alpha'}]\nonumber\\
&+\frac{\gamma_{\phi}}{2}\sum_i\mathcal{L}[L_{b_i,b_i}]\label{Atomic-Eq.Mot}
\end{align}
where $L_{\alpha,\alpha'}=\ket{\alpha}\bra{\alpha'}$ and $L_{b_i,b_i}=\ket{b_i}\bra{b_i}$. 
Final two terms of Eq.~(\ref{Atomic-Eq.Mot}) corresponds to $\mathcal{L}_A[\rho_A]$ with $\alpha'\neq\alpha$ and $i=1,..,N$. Here, $\alpha=\{a,b_1,...,b_N\}$ and  $\gamma_{\alpha,\alpha'}$ is taken equal to $\gamma$ for simplicity. Note that the atomic part of the master equation is for the case of pure dephasing $\&$ relaxation and we follow the usual assumption of micromaser theory~\cite{filipowicz_theory_1986} that cavity decay and atomic dynamics can be separately treated. When atom is inside the cavity, decay is not included.  

The equation of motion of the $i^{th}$ term of the first summation of Eq.~(\ref{9EqMot}) for a single atom is
\begin{align}
\dot{\rho}_{b_i n+1,an}&=-ig\sqrt{n+1}e^{i\Omega t}\{\rho_{an,an}\nonumber\\
&-(\rho_{b_i n+1,b_1 n+1}+..+\rho_{b_i n+1,b_N n+1})\}.\label{Eq10}
\end{align}
Here, we have neglected the the matrix element $\bra{b_i n+1}\mathcal{L}_f[\rho]\ket{an}$ in accord with the assumptions indicated above. 
The equation of motion for $\rho_{b_i n,a n-1}$ which is the $i^{th}$ term of second summation of (\ref{9EqMot}) could be obtained by simply replacing $n\rightarrow n-1$ in Eq.~(\ref{Eq10}), that is
\begin{align}
\dot{\rho}_{b_i n,an-1}&=-ig\sqrt{n}e^{i\Omega t}\{\rho_{an-1,an-1}\nonumber\\
&-(\rho_{b_i n,b_1 n}+..+\rho_{b_i n,b_N n})\}\label{Eq.11}
\end{align}
Therefore we obtain $ \rho_{b_i n+1,an}$ and $ \rho_{b_i n,an-1}$ terms by integrating Eq.s~(\ref{Eq10} and \ref{Eq.11}) formally in the following form,

\begin{eqnarray}
\rho_{b_i n+1,an}^k=-ig\sqrt{n+1}\int_{t_{k_0}}^t dt' e^{(i\omega_{a b_i}-\gamma)(t-t')}e^{i\Omega t'}\nonumber\\
\times\{\rho_{an,an}^k-(\rho_{b_i n+1,b_1 n+1}^k+..+\rho_{b_i n+1,b_N n+1}^k)\}\nonumber\\\label{Eq12}
\end{eqnarray}

\begin{eqnarray}
\rho_{b_i n,an-1}^k=-ig\sqrt{n}\int_{t_{k_0}}^t dt' e^{(i\omega_{a b_i}-\gamma)(t-t')}e^{i\Omega t'}\nonumber\\
\times\{\rho_{an-1,an-1}^k-(\rho_{b_i n,b_1 n}^k+..+\rho_{b_i n,b_N n}^k)\}\nonumber\\\label{Eq13}
\end{eqnarray}
The terms of Eq.s~(\ref{Eq12} and \ref{Eq13}) can be factorized to atomic and field density matrices for zeroth order solution in g. For instance, 
\begin{align}
\rho_{\alpha n,\alpha n}^{k_0}(t',t_{k_0})=\rho_{\alpha,\alpha}^{k_0}(t',t_{k_0})\rho_{n,n}(t') ,\nonumber\\
\rho_{\alpha n+1,\alpha n+1}^{k_0}(t',t_{k_0})=\rho_{\alpha,\alpha}^{k_0}(t',t_{k_0})\rho_{n+1,n+1}(t'),\nonumber\\
\end{align}
\begin{eqnarray}
\rho_{b_i n,b_j n}^{k_0}(t',t_{k_0})=\rho_{b_i,b_j}^{k_0}(t',t_{k_0})\rho_{n,n}(t'),\nonumber\\
\rho_{b_i n+1,b_j n+1}^{k_0}(t',t_{k_0})=\rho_{b_i,b_j}^{k_0}(t',t_{k_0})\rho_{n+1,n+1}(t')\nonumber\\\label{Eq14}
\end{eqnarray}
Here, $\rho_{\alpha,\alpha}^{k_0}$, $\rho_{b_i,b_j}^{k_0}$ which are the initial atomic density matrix elements, obey the respective atomic equations of motion
\begin{eqnarray}
\dot{\rho}_{\alpha,\alpha}^{k_0}(t',t_{k_0})=-\gamma\rho_{\alpha,\alpha}^{k_0}(t',t_{k_0}),\nonumber\\
\dot{\rho}_{b_i,b_j}^{k_0}(t',t_{k_0})=-(i\omega_{b_i b_j}+\gamma+\gamma_{\phi})\rho_{b_i b_j}^{k_0}(t',t_{k_0})\nonumber\\\label{Eq15}
\end{eqnarray}
in which the solutions are
\begin{eqnarray}
\rho_{\alpha,\alpha}^{k_0}=e^{-\gamma(t-t_{k_0})}\rho_{\alpha,\alpha}^{k_0}(t_{k_0},t_{k_0}),\nonumber\\
\rho_{b_i,b_j}^{k_0}=e^{-(i\omega_{b_i b_j}+\bar{\gamma})(t-t_{k_0})}\rho_{b_i,b_j}^{k_0}(t_{k_0},t_{k_0})\label{Eq16}
\end{eqnarray}
where $\bar{\gamma}=\gamma+\gamma_{\phi}$. Eq.s (\ref{Eq15}) and (\ref{Eq16}) imply that excited and ground state levels decay to a lower level. The off-diagonal elements of the atomic density matrix are equal to $\rho_{b_i,b_j}^{k_0}(t_{k_0},t_{k_0})=|\rho_{b_i,b_j}^{0}|e^{i\phi_{ij}}$. The $e^{i\phi_{ij}}$ is assigned with the coherence preparation. 
Thus, we can express Eq.s~(\ref{Eq12} and \ref{Eq13}) by using zeroth order atomic Eq.s~(\ref{Eq14}-\ref{Eq16}) to find first order solutions in g,  
 
\begin{eqnarray}
\rho_{b_i n+1,an}^k=-ig\sqrt{n+1}\int_{t_{k_0}}^t dt' e^{(i\omega_{a b_i}-\gamma)(t-t')}e^{i\Omega t'}\nonumber\\
\times\{e^{-\gamma(t'-t_{k_0})}\rho_{aa}\rho_{nn}-e^{-\gamma(t'-t_{k_0})}\rho_{b_i b_i}\rho_{n+1 n+1}\nonumber\\
 -\sum_{i\neq j}e^{-(i\omega_{b_i b_j}+\bar{\gamma})(t-t_{k_0})}|\rho_{b_i,b_j}^{0}|e^{i\phi_{ij}}\rho_{n+1 n+1}\}.\nonumber\\\label{Eq17}
\end{eqnarray}

\begin{widetext}

Likewise,
\begin{align}
\rho_{b_i n,a n-1}^k&=-ig\sqrt{n}\int_{t_{k_0}}^t dt' e^{(i\omega_{a b_i}-\gamma)(t-t')}e^{i\Omega t'}\{e^{-\gamma(t'-t_{k_0})}\rho_{aa}^0\rho_{n-1 n-1}-e^{-\gamma(t'-t_{k_0})}\rho_{b_i b_i}^0\rho_{n n}\nonumber\\
&-\sum_{i\neq j}e^{-(i\omega_{b_i b_j}+\bar{\gamma})(t-t_{k_0})}|\rho_{b_i,b_j}^{0}|e^{i\phi_{ij}}\rho_{n n}\}.\nonumber\\\label{Eq18}
\end{align}
Putting all these results into Eq.~(\ref{9EqMot}) we have the field equation of motion,

\begin{align}
\dot{\rho}_{n n}&=-g^2\sum_k\int_{t_{k_0}}^t dt'\{e^{-i\Omega (t- t')}[(n+1)\sum_{i=1}^{N}e^{(i\omega_{a b_i}-\gamma)(t-t')}
(e^{-\gamma(t'-t_{k_0})}\rho_{aa}^0\rho_{n n}-e^{-\gamma(t'-t_{k_0})}\rho_{b_i b_i}^0\rho_{n+1 n+1}\nonumber\\
&-\sum_{i\neq j}e^{-(i\omega_{b_i b_j})+\gamma)(t'-t_{k_0})}|\rho_{b_i,b_j}^{0}|e^{i\phi_{ij}}\rho_{n+1 n+1})-n\sum_{i=1}^{N}e^{(i\omega_{a b_i}-\gamma)(t-t')}(e^{-\gamma(t'-t_{k_0})}\rho_{aa}^0\rho_{n-1 n-1}-e^{-\gamma(t'-t_{k_0})}\rho_{b_i b_i}^0\rho_{n n}\nonumber\\
&-\sum_{i\neq j}e^{(i\omega_{b_i b_j}+\bar{\gamma})(t'-t_{k_0})}|\rho_{b_i,b_j}^{0}| e^{i\phi_{ij}}\rho_{n n})]+c.c.\}\nonumber\\\label{Eq19}
\end{align}
Before proceeding, we replace the summation over number of injected atoms by integration over injection time as $\sum_k\rightarrow r\int_{-\infty}^{t} d k_0$ where $r$ is the injection rate. We also define $\Delta_i=\omega_{a b_i}-\Omega$ where $\omega_{a b_i}=\omega_a-\omega_{b_i}$. Then,  
\begin{align}
\dot{\rho}_{n n}&=-rg^2\int_{-\infty}^t dt_{k_0}\int_{t_{k_0}}^t dt'\Big\{\Big[\sum_{i=1}^{N}e^{(i\Delta_i-\gamma)(t'-t_{k_0})}\rho_{aa}^0((n+1)\rho_{nn}-n\rho_{n-1 n-1})\nonumber\\
&-(\sum_{i=1}^{N}e^{(i\Delta_i-\gamma)(t-t')}e^{-\gamma(t'-t_{k_0})}\rho_{b_i b_i}^0)
((n+1)\rho_{n+1 n+1}-n\rho_{nn})\nonumber\\
&-(\sum_{i<j}(e^{(i\Delta_i-\gamma)(t-t')} e^{-(i\omega_{b_i b_j}+\bar{\gamma})(t'-t_{k_0})}|\rho_{b_i,b_j}^{0}|e^{i\phi_{ij}}
+e^{(i\Delta_j-\gamma)(t-t')}e^{-(-i\omega_{b_i b_j}+\bar{\gamma})(t'-t_{k_0})}|\rho_{b_i,b_j}^{0}|e^{-i\phi_{ij}})\nonumber\\
&\times((n+1)\rho_{n+1 n+1}-n\rho_{nn})\Big]+c.c\Big\}+\mathcal{L}_f[\rho]_{nn}.\label{Eq20}
\end{align}
Note that $\rho_{b_j b_i}=e^{-(-i\omega_{b_i b_j}+\bar{\gamma})(t-t_{k_0})}|\rho_{b_i,b_j}^{0}|e^{-i\phi_{ij}}$ while $\rho_{b_i b_j}=e^{-(i\omega_{b_i b_j}+\bar{\gamma})(t-t_{k_0})}|\rho_{b_i,b_j}^{0}|e^{i\phi_{ij}}$. Evaluating the integrals in  (\ref{Eq20}) over $t'$ and $t_{k_0}$ after changing integration order as $\int_{-\infty}^t dt_{k_0}\int_{t_{k_0}}^t dt'=\int_{-\infty}^t dt'
\int_{-\infty}^{t'} dt_{k_0}$ we have

\begin{align}
\dot{\rho}_{n n}&=-rg^2\Bigg\{\frac{1}{\gamma}\Big(\sum_{i=1}^{N}\frac{1}{-i\Delta_i+\gamma}\Big)\rho_{aa}^0\big[(n+1)\rho_{nn}-n\rho_{n-1 n-1}\big]-\Big[\frac{1}{\gamma}\Big(\sum_{i=1}^{N}\frac{1}{-i\Delta_i+\gamma}\Big)\rho_{b_i b_i}^0\nonumber\\
&+\sum_{i<j}\Big(\frac{1}{(-i\Delta_i+\gamma)}\frac{1}{(i\omega_{b_i b_j}+\bar{\gamma})}e^{i\phi_{ij}}+\frac{1}{(-i\Delta_j+\gamma)}\frac{1}{(-i\omega_{b_i b_j}+\bar{\gamma})}e^{-i\phi_{ij}}\Big)|\rho_{b_i,b_j}^{0}|\Big]\nonumber\\
&\times\big[(n+1)\rho_{n+1 n+1}-n\rho_{nn}\big]+c.c.\Bigg\}+\mathcal{L}_f[\rho]_{nn}\label{Eq21}
\end{align}
We proceed by summing each term with their respective complex conjugates, some arrangements then,

\begin{align}
\dot{\rho}_{n n}&=-rg^2\Bigg\{\frac{1}{\gamma}\Big(\sum_{i=1}^{N}\frac{2\gamma}{\Delta_i^2+\gamma^2}\Big)\rho_{aa}^0\big[(n+1)\rho_{nn}-n\rho_{n-1 n-1}\big]-\Big[\frac{1}{\gamma}\Big(\sum_{i=1}^{N}\frac{2\gamma}{\Delta_i^2+\gamma^2}\Big)\rho_{b_i b_i}^0\nonumber\\
&+\sum_{i<j}\Big[\Big(\frac{2\cos\phi_{ij}(\Delta_i\omega_{b_i b_j}+\gamma\bar{\gamma})+2\sin\phi_{ij}(\omega_{b_i b_j}-\Delta_i\bar{\gamma})}{(\Delta_i^2+\gamma^2)(\omega_{b_i b_j}^2+\bar{\gamma}^2)}+\frac{2\cos\phi_{ij}(\gamma\bar{\gamma}-\Delta_j\omega_{b_i b_j})+2\sin\phi_{ij}(\omega_{b_i b_j}+\Delta_j\bar{\gamma})}{(\Delta_j^2+\gamma^2)(\omega_{b_i b_j}^2+\bar{\gamma}^2)}\Big)|\rho_{b_i b_j}^0|\Big]\nonumber\\
&\times\big[(n+1)\rho_{n+1 n+1}-n\rho_{nn}\big]\Bigg\}\label{Eq22}
\end{align}

and by using Eq.~(\ref{Louvil-Rho}), finally we have,

\begin{align}
\dot{\rho}_{n n}&=-R\Big\{K_a\rho_{aa}\big[(n+1)\rho_{nn}-n\rho_{n-1 n-1}\big]+(R_{g_0}+R_{g_c})\times\big[n\rho_{nn}-(n+1)\rho_{n+1 n+1}\big]\Big\}\nonumber\\
&+\kappa\left\lbrace(n+1)\rho_{n+1,n+1}-n\rho_{nn}\right\rbrace\label{Eq23}
\end{align}
where
\begin{align}
K_a=\sum_{i=1}^N \frac{2}{\Delta_i^2+\gamma^2} ,\quad  R_{g_0}=\sum_{i=1}^N K_{b_i}\rho_{b_ib_i}^0,
 R_{g_c}=\sum_{i<j}^S K_{ij}^{\phi_{ij}}|\rho_{b_i b_j}^0|,\quad K_{b_i}=\frac{2}{\Delta_i^2+\gamma^2},\label{Eq24}
\end{align}

\begin{align}
K_{ij}^{\phi_{ij}}=\frac{2\cos\phi_{ij}(\Delta_i\omega_{b_i b_j}+\gamma\bar{\gamma})+2\sin\phi_{ij}(\omega_{b_i b_j}-\Delta_i\bar{\gamma})}{(\Delta_i^2+\gamma^2)(\omega_{b_i b_j}^2+\bar{\gamma}^2)}
+\frac{2\cos\phi_{ij}(\gamma\bar{\gamma}-\Delta_j\omega_{b_i b_j})+2\sin\phi_{ij}(\omega_{b_i b_j}+\Delta_j\bar{\gamma})}{(\Delta_j^2+\gamma^2)(\omega_{b_i b_j}^2+\bar{\gamma}^2)},\label{Eq25}
\end{align}
$\Delta_{i,j}=\omega_{a b_{i,j}}-\Omega$, $\omega_{a b_{i,j}}=\omega_a-\omega_{b_{i,j}}$ and $R=rg^2$. Note that $ R_{g_0}$ has $N$ number of terms and $ R_{g_c}$ has $S=N(N-1)/2$ number of terms in the summation. 
Since we seek the solutions in the steady state, we obtain the steady state photon number $\bar{n}_{\phi}$ by solving $\dot{\bar{n}}_{\phi}=0$ where $\dot{\bar{n}}=\sum_n n\dot{\rho}_{nn}$ and we write $\dot{\bar{n}}_{\phi}$ by using previously obtained $\dot{\rho}_{n n}$ as
\begin{align}
\dot{\bar{n}}&=-RK_a\rho_{aa}\sum_n n(n+1)\rho_{nn}+RK_a\rho_{aa}\sum_n n^2\rho_{n-1,n-1}
-RR_{g_0}\sum_n n^2\rho_{nn}+RR_{g_0}\sum_n n(n+1)\rho_{n+1,n+1}\nonumber\\
&-RR_{g_c}\sum_n n^2\rho_{nn}+RR_{g_c}\sum_n n(n+1)\rho_{n+1,n+1}
+\kappa\sum_n n(n+1)\rho_{n+1,n+1}-\kappa\sum_n n^2\rho_{nn}.\label{Eq26}
\end{align}
\end{widetext}

Then, we insert $n\rightarrow n-1$ for and $n\rightarrow n+1$ for $\rho_{n+1 n+1}$ and $\rho_{n-1 n-1}$ terms respectively so that we get,

\begin{align}
\dot{\bar{n}}&=RK_a\rho_{aa}\sum_n (n+1)\rho_{nn}-RR_{g_0}\sum_n n\rho_{nn}\nonumber\\
&-RR_{g_c}\sum_n \rho_{nn}\nonumber\\
&=RK_a\rho_{aa}(\bar{n_{\phi}}+1)-R\bar{n_{\phi}}(R_{g_0}+R_{g_c})-\kappa\bar{n}_{\phi}\label{Eq27}
\end{align}
Solving  $\dot{\bar{n}}_{\phi}=0$,  we have

\begin{align}
\bar{n}_{\phi}&=\frac{K_a\rho_{aa}}{R_{g_0}+R_{g_c}+\frac{\kappa}{R}-K_a\rho_{aa}}\nonumber\\
&=\frac{1}{\frac{R_{g_0}}{K_a\rho_{aa}}+\frac{R_{g_c}}{K_a\rho_{aa}}+\frac{\kappa}{RK_a\rho_{aa}}-1}.\label{Eq28}
\end{align}
We write the final form of the steady state photon number after some arrangements,

\begin{eqnarray}
\bar{n}_{\phi}=\frac{\bar{n}_{\kappa}}{1+\bar{n}_{\kappa}\frac{R_{g_c}}{K_a\rho_{aa}}},\label{Eq29}
\end{eqnarray}
where
\begin{eqnarray}
\bar{n}_{\kappa}=\frac{\bar{n}}{1+\bar{n}\frac{\kappa}{RK_a\rho_{aa}}}.  \label{Eq30}       
\end{eqnarray}
Here, $\bar{n}_{\kappa}$ is the average photon number in the  absence of atomic coherence in terms of average photon number $\bar{n}=1/(R_{g_0}/K_a\rho_{aa}-1)$ which is the average photon number in the absence of atomic coherence and in the absence of cavity decay $\kappa$.  

Finally we look at the degenerate ground state  case ($\omega_{a b_{i,j}}=0$, $\Delta_{i,j}=0$). 
In this case, $\bar{n}$ can be simplified to $\bar{n}=P_e/(Pg-Pe)$ where $P_e=\rho_{aa}$, $P_g=\rho_{b_i b_i}$ for any $i$. The simplified forms of other parameters are

\begin{align}
K_a=\frac{2N}{\gamma^2}, \quad R_{g_0}=\frac{2NP_g}{\gamma^2}, \quad Rg_c=\frac{2N(N-1)\cos\phi\lambda}{\gamma\bar{\gamma}}\nonumber\\
\end{align}
for $\theta=\pi$. Analytical decoherence term for degenerate case can be identified in $R_{g_c}$ expression such that 

\begin{eqnarray}
\xi= (1+\frac{\gamma_{\phi}}{\gamma})^{-1}\cong e^{-\gamma_{\phi}/\gamma}
\end{eqnarray}
for $\gamma_{\phi}\ll\gamma$.



%


\end{document}